\documentclass[fleqn,twoside]{article}
\usepackage{espcrc2}
\usepackage{amsmath}
\usepackage{amssymb}
\title{The Mysteries of the Highest Energy Cosmic Rays}
\author{S. Kovesi-Domokos\address{Department  
of Physics and Astronomy\\ The
Johns Hopkins University, Baltimore, MD}\thanks{Speaker. e-mail: {\tt skd@jhu.edu}\hspace{1em} Invited talk given at XIII ISVHECRI, Pylos, Greece.}
and
G. Domokos\addressmark\thanks{e-mail: {\tt Gabor.Domokos@jhu.edu}}}
\begin{document}
\newcommand{\lappr}{\mbox{$\stackrel{<}{\sim}$}} 
\newcommand{\gappr}{\mbox{$\stackrel{>}{\sim}$}} 

\begin{abstract}
We give a summary of the theoretical attempts to make sense of the observational data
concerning the highest energy cosmic rays, $E > 3\times 10^{19}$eV.

\end{abstract}

\maketitle

\section{Preamble}

It was about forty years ago that Linsley~\cite{linsley} presented 
an incredulous physics community with the $\sim 10^{20}$~eV  event
observed at the Volcano Ranch array. Today ground arrays and fluorescence
detectors have collected over a hundred events with energies above 
$3\times 10^{19}$ eV,  and a dozen or so at or above $ 10^{20}$ eV . It is worth remembering  
that the $\sim2$~TeV CM energies attained at the Tevatron are at the {\em knee} of the 
cosmic ray spectrum.  LHC will operate at an equivalent fixed target energy 
falling approximately halfway between the {\em knee} and the {\em ankle},
around the so-called ``{\em second knee}''  region. Thus, above the {\em second knee}, 
the first interaction  in  an extensive air shower involves  physics untested
by the well-controlled experimental environments of accelerators.\\

For many years, the appearance of the GZK feature~\cite{GZK}
at the highest energies (i.e. above $2 \times 10^{19}$ eV)  was taken for granted.  
The processes, $p + \gamma_\text{CMB} \rightarrow \pi + X$, and, of lesser importance,
$p+ \gamma_\text{CMB}\rightarrow e^+ e^- + X$, are well-studied low energy 
processes in the CM. A proton of energy $10^{20}$~eV or above has a mean free path of 
7--8 Mpc in the cosmic microwave background (CMB) radiation. With each interaction event,
the proton loses   20--25\% of its energy; after only a few interactions its energy drops 
under the threshold of the pion photoproduction process. 
The expected precipitous drop in the
incoming cosmic ray flux at $5\times 10^{19}- 8\times 10^{19}$ eV (GZK cutoff) and 
the pileup~\cite{schramm}
(a bump in the spectrum) just before the drop\footnote{This is a consequence 
of the pion photoproduction by
the above threshold protons or nuclei.} are  called  ``the GZK feature''. 
The GZK sphere 
approximately defines the cosmic neighborhood of sources from which we expect
the arrival of  extreme high energy cosmic rays ($E\approx 10^{20}$~eV and up). 
We usually put this particle
horizon at 50--70 Mpc. However, the GZK radius could be considerably less as 
demonstrated by Stanev {\em et al.}~\cite{stanev1}. Scattering in  extragalactic
random  magnetic fields makes the real distance travelled by the particle longer than the 
source-Earth distance; more energy loss processes can take place as a result.
Heavier nuclei with similar energies also suffer energy loss through photodisintegration
on the cosmic microwave and on the intergalactic
infrared background radiations~\cite{stecker}.
 For $^{56}\text{Fe}$,
for example, the decrease of the energy loss time is even more abrupt than for protons,
though it starts at a somewhat higher energy. \\
  
We will  summarize the most important, and often controversial, observational data.
Then we present the theories formulated to explain certain coherent subsets
of the data.  
\section{Observational Data}

What follows is a very brief
summary of the most important features of the
extensive air shower and EGRET data from a theorist's point of view. Only ultra high
energy cosmic rays (UHECR) with energies above $3\times 10^{19}$ eV are considered.

\subsection{Spectrum }
\vspace{2mm}
At and above $10^{20}$~eV, there are 15 events: 11 observed by 
AGASA~\cite{agasa} and 
2 by HiRes~\cite{hires} and one each by Fly's Eye~\cite{fly},  and Yakutsk~\cite{yakutsk}.\\

There is a substantial  disagreement between AGASA and HiRes  at this time.  
The HiRes data were successfully fitted~\cite{hires2}
with a two-component model~\cite{waxman} including primaries with galactic and  extragalactic origins.
The Fly's Eye data suggest a transition from heavy to light (proton) primaries (at energies from
$10^{17}$ to $10^{19}$~eV), which can be interpreted as a sign of the passage from  galactic 
to extragalactic sources.The  extragalactic source model of  
Berezinsky {\em et al.}~\cite{berezinsky},
which takes into account both uniform source distribution and a maximal energy at $10^{21}$~eV,
yields not only the GZK feature
but also other details of the spectrum, as a result of considering both energy loss 
processes. The large number of events above the GZK cutoff (mostly observed by AGASA)
remains unexplained by the two-component models\footnote{Since for both AGASA and HiRes there
are systematic errors in the energy assignment, it is difficult to judge how great
the disagreement between them actually is.}. \\

It should be emphasized that even the two largest experiments, AGASA and HiRes, have too few
events at the highest energies; the discrepancy between them has low statistical 
significance as a result~\cite{demarco}.

\subsection{Composition}
\vspace{2mm}
 As mentioned above,  fluorescence detectors
provide some hints for a Fe $\rightarrow$ p transition.
However, it is important to note that the mass
composition results are different for each of the various methods and model calculations used\cite{watson}.
In addition, the shower energy estimates are sensitive to the assumption about the nature of the primary.
To quote A.A.~Watson~({\em loc. cit.}), ``knowledge about the mass of the 
primary cosmic rays
at energies above $10^{17}$ eV is rudimentary.''
As far as photon primaries are concerned, it is very unlikely that the primary of
the ``gold-plated'' $3\times 10^{20}$~eV Fly's Eye event was a photon~\cite{halzen}.
Around $10^{19}$~eV, the photon flux is less than 40\% of the baryonic component according to both
Haverah Park~\cite{haverah} and AGASA~\cite{agasa2}. By searching for deeply penetrating
 horizontal air showers, AGASA
investigated the UHE neutrino flux for energies $> 3\times 10^{17}$ eV~\cite{yoshida}.
No significant enhancement above the expected hadron-induced flux was reported.

\subsection{Directionality}
\vspace{2mm}
 For the energy range we are interested in, the arrival direction
distribution is remarkably isotropic. This could indicate a large number of
weak or  distant sources.  AGASA data show small-scale clustering~\cite{takeda}
(4 doublets and one triplet within a separation angle of $2.5^{\text{o}}$) above $4\times10^{19}$eV
(57 events). HiRes in stereo mode above $10^{19}$eV (162 events) claims to see no small-scale anistropy
in the arrival direction distribution~\cite{finley}. Recently, it was found~\cite{westerhoff}
that the statistical significance here is also very low for 
both the small scale clustering in the AGASA data set and the apparent discrepancy between AGASA
and HiRes.\\

Nevertheless, despite the paucity of
data points, statistical analyses have attempted  to discover 
possible angular correlations between the arrival directions of the extremely high energy
cosmic rays (EHECR), and to match these in turn with high energy source canditates. 
The enticement to carry  out an analysis of even the small available
sample is obvious. Significant angular correlation 
in the arrival directions of the EHECR could be the a smoking gun for the
existence of discrete compact (repeating) sources.
Using the AGASA data of $E>4\times10^{19}$ eV~\footnote{At $E=4\times10^{19}$eV protons have a loss
length of $\sim 1000 Mpc$.}, the presence of clustering can be used 
to constrain the number of sources  in the nearby Universe for proton primaries\cite{demarco2}
to be $~10^{-5} Mpc^{-3}$ with large uncertainties. Tinyakov and Tkachev~\cite{Russians} claimed to
have identified  BL Lacertae objects as sources generating the anisotropies in the AGASA data.
Evans {\em et al.}~\cite{Subir} and W.~Burgett {\em et al.}~\cite{burgett} questioned
this conclusion, however\footnote{Burgett and 
O'Malley partition the AGASA sample into three statistically uncorrelated groups by energy:
$E < 5 \times 10^{19}$ eV, $5 \times 10^{19}\leq E < 8 \times 10^{19}$ eV, and $E > 8 \times
10^{19}$ eV. They find the 
middle group to be preferentially aligned with the Supergalactic equatorial plane,while the 
other two groups are statistically consistent with isotropic distributions.}.\\ 

Most authors working in the field agree that more data from HiRes and from  larger
experiments like the Southern (and hopefully the Northern) Pierre Auger Observatory, EUSO,
the Telescope Array, and OWL are necessary to draw 
solid conclusions about clustering and the presence or absence of the GZK cutoff discrepancies. 
Until then the BL Lac alignment is merely suggested by the need for objects accelerating particles
to exceptional energies, rather than being a firmly established consequence of
the observational data.

\subsection{Extragalactic Diffuse Gamma Rays}
\vspace{2mm} 
  Another piece  of important data is the extragalactic diffuse $\gamma$
ray background (EGRB).  It is generally believed that
unresolved AGN, galaxy clusters, distant gamma ray bursts, {\em etc.} are the main sources.
However, EGRB would  get a significant contribution from
the decay of  topological objects, relic particles {\em etc.},
which are used by some authors to explain trans-GZK cosmic ray events.
Comparing that contribution to the
measured EGRB, one may be able to place upper limits on, or eventually
exclude, some of those schemes. Any electromagnetic (EM) energy introduced to the 
EM background radiation (dominated by CMB) produces an EM cascade if the energy is 
$\gtrsim 10^{15} \text{eV}/(1+z)$ {\em i.e.} the pair production ``threshold'' on the CMB at redshift
z.  The cascade by pair production and inverse Compton scattering degenerates the injected energy
to under $\sim 10^{15}$ eV in a couple of Mpc and produces a power-law distribution
of  GeV-range gamma rays.\\

To obtain the EGRB, one must construct a model of the {\em galactic} $\gamma$ ray
background, subtract it from the observational data, and assign the remainder
to the EGRB. This is a complicated and model-dependent process.
The EGRB was  first discovered  with SAS-II~\cite{thompson} and then confirmed  
by EGRET~\cite{sreekumar}. A power law energy spectrum ($E^{(-(2.1 \pm 0.03))}$ in 
the range 30 MeV $\lesssim E \lesssim $100 GeV)
with no anisotropy was found.
Strong {\em et al.}~\cite{strong}, in a recent reworking of the EGRET data, substantially improved
the determination of the diffuse galactic $\gamma$ ray flux. They conclude  that in previous analyses 
the galactic contribution was underestimated; the EGRB must be revised 
{\em downward} as a result. The new spectrum is not consistent with a power law.
One may infer tentatively that the EGRB arises from a number of different physical processes.\\

At higher energies ($ 20\text{TeV}\lesssim E \lesssim 500\text{TeV}$), 
an upper limit on the EGRB can be obtained by estimating the ratio of 
photon-induced to hadron-induced extensive air showers.
The GRAPES collaboration~\cite{grape} obtained an 
upper limit of about $3\times 10^{-4}$ for the fraction of
photon-induced (muon-poor) air showers.

\section{Theoretical Models}

  The presence or absence of the GZK
feature determines whether it is enough to rely on conventional astrophysics
to explain the spectrum, composition, and directionality encrypted in  the data, or if
physics beyond the Standard Model is needed \cite{reviews}.

\subsection{Traditional Astrophysics}

\vspace{2mm}

These models are also called {\em bottom-up scenarios}. At $10^{19}$~eV and above, the extragalactic
component must dominate the energy spectrum.  Even if  there are some objects
(neutron stars, galactic wind, {\em etc}) in our
galaxy which could
possibly accelerate UHECR, it would be  impossible for them to produce the observed
isotropic distribution. 
 For order of magnitude estimates, the 
community generally uses Hillas' formula. Whether the acceleration takes place 
electromagnetically or with the help of diffusive shock acceleration (first order Fermi process),
the magnetic field must be strong enough to keep the particles in the
 region of acceleration, thus, approximately, 
\[ R_\text{accelerator} \gtrsim R_\text{gyromagnetic} \approx \frac{E}{Ze B},\]
where the charge of the accelerated particle is $Ze$. From this,
\[E_{max} \approx Ze B R_{accelerator}.\] 
We are looking for a maximal energy of about $10^{21} - 10^{22}$~eV. 
The enormous energy of the primary singles out exceptional
astrophysical sources: radio lobes of powerful radio galaxies, jets of active galactic nuclei,
gamma ray bursters, and magnetars. There are only a few of these possible accleration sites inside the
GZK radius for charged hadrons. (For example, the radio galaxy M87, in the center of the Virgo cluster,
was investigated repeatedly as a candidate\cite{protheroe2}. It could provide EHECR at Earth only
if the extragalactic magnetic fields on the way here have some specific configuration.) 
Consequently, it is unlikely that the  GZK cutoff could be completely absent in
the spectrum. The accelerators are far beyond the pion photoproduction distance and the
random, poorly known, extragalactic magnetic fields may
 cause deflection of charged particles.  
In fact, the relevant sources are at cosmological distances. Nevertheless there are
small scale anisotropies and correlations with certain sources. These observations
indicate that the cosmic traveler may be a stable, neutral particle
which interacts very weakly with the cosmic background radiation fields. This particle
could be a neutrino.\\

It is generally accepted that the dominant form of acceleration of cosmic rays at the
sources is by first order Fermi acceleration at astrophysical shocks. A recent development 
in the theory of shock acceleration is the introduction of the back reaction of the accelerated particles
to the shock modifying it strongly\cite{jones}. Among other results the spectrum is claimed
to be flatter than in the linear theory; thus, the number of arriving trans-GZK primaries
could be higher. 
 
\subsection{ Scenarios Beyond the Standard Model}

\vspace{2mm}

\subsubsection{Z-bursts}
\vspace{1mm}
 The Z-burst model\cite{weiler} needs only the inclusion of the neutrino mass
in the Standard Model (SM) for its attempt to explain the trans-GZK events. Among SM particles
only the neutrino can handle cosmic distances without dramatic absorption. The assertion is that
extreme energy neutrinos
 \[E_{res} = M_{Z}^{2}/2m_{\nu} = 4\times10^{21}(m_{\nu}(eV))^{-1}eV \]
  \[= 4\times(m_{\nu}(eV))^{-1}ZeV\]
 forming $Z^0$ bosons resonantly with
antineutrinos of the relic 1.9 $^{\text{o}}\text{K}$ thermal neutrino background could produce enough
protons well inside the GZK radius to account for the EECR. On average a $Z^0$ produces  
$\sim2$ nucleons, $\sim20$ photons and $\sim50$ neutrinos.  Although the Z-burst model is very
 economical,
but it is likely to have fatal problems both 
because of the required energy  at the source and because of the  extreme energy neutrino flux
 required  to produce enough of the trans-GZK events~\cite{domokos}. The new experimental limit
by WMAP, 2dF and SDSS on the active neutrino mass ($\sum m_{\nu i} \lesssim 0.72$ eV) combined 
with the neutrino oscillation results gives the maximal possible neutrino mass 
$m_{\nu}\approx 0.24$ eV~\cite{bhattar}. This puts the  energy  of the neutrino for 
the resonant production at $\sim 10$ ZeV. If these neutrinos
are secondaries of pions, the  protons in the source must be accelerated to $\sim 10^3$ ZeV. 
No known astrophysical object is capable of accelerating protons to such high 
energies~\cite{kalashev}. A large enough flux
by ordinary astrophysical sources (or by beyond-the Standard Model objects~\cite{sigl})
to produce a substantial part  of trans-GZK events  would pump so much EM
energy into the universe that the EGRET limits would be violated many times over.\\

An overdensity of neutrinos
in a halo of our galaxy or in that of the Local Group could help to reduce 
the flux requirement. However, these neutrino masses are too small to allow any clustering of
relevance to allow the reduction of the incoming neutrino flux~\cite{fodor}, \cite{semikoz}.
For a conservative maximal neutrino mass, $m_{\nu}\approx0.33$ eV, the neutrino flux required
by the Z-burst scenario
is already in marginal conflict with the upper limit given by the FORTE~\cite{forte} experiment and
in violation of the GLUE~\cite{glue} limit. A lower mass makes the
problem more severe~\cite{semikoz}. This limit on the Z-burst model is valid even when the distant
sources emit only neutrinos. \\

\subsubsection{Strongly Interacting Neutrinos}
\vspace{1mm}
The problem of energy loss during the propagation at cosmological distances
would not arise if the trans-GZK primaries
were neutrinos and interacted strongly at very high energies~\cite{berzat}, causing proton-like
air showers. Both the energy and flux reqirements are much lower here than in the Z-burst model.
In this case the cosmogenic neutrino flux ({\em i.e.} secondaries from the pion photoproduction
of extreme energy protons on the CMB) could account for the trans-GZK events~\cite{zfodor},
since all incoming neutrinos produce an air shower. For the protons
the injection spectrum is simple, power law.
  The onset of strong interaction must
result in a very rapidly rising cross section; a slow rise would give a large number of horizontal 
showers, which are not seen by either AGASA or HiRes. For example, in theories of higher than 4 dimensions,
Kaluza-Klein graviton exchange
gives a power law increase, which is not only too slow, but strong interaction cross sections cannot
be reached until energies $\approx 10^{20}$ eV~\cite{anch}.\\

 The following two schemes also assume some higher 
dimensional (string inspired) theory.  The neutrino-nucleon cross section begins to grow at a certain
energy, in essence due to the unification of all interactions at
a scale of the order of a few tens of TeV or a few TeV in the CMS,
producing lepto-quark resonances~\cite{gabor},
mini black holes~\cite{luis} or ``p-branes''~\cite{ahn}.
 The asymptotic ($s\rightarrow\infty$) equivalence of the resonance and mini black hole schemes
was recently demonstrated~\cite{zsuzsa}. The {\em onset} of the new interaction is very different.
In the ``semiclassical black hole'' picture, it is assumed that the cross section is proportional to
the square of the higher dimensional (d=10 ?) Schwarzschild radius. Combined with the assumption 
that $m_{\text{bh}}\propto\sqrt{s}$, it leads to an unacceptably slow rise of the cross section.
The published data on near horizontal showers by AGASA and Fly's Eye exclude that. By contrast,
with a string scale of the order 70~TeV one is far below the semiclassical region~\cite{zsuzsanna}.  
The growth of the cross section is governed by the exponentially increasing level density in the
transition region (and only in the transition region) from the SM to the string regime. In this scheme
few near horizontal showers are produced, as a result.\\

Another possibility,  suggested 
to achieve strong cross sections, is  the inclusion of
instanton induced processes. There is a detailed discussion and further references
on the instanton enhanced cross section in \cite{zfodor}.\\ 

It is remarkable that the cosmogenic neutrino beam, together with a strongly interacting neutrino
with  sudden onset of large cross section could give a solution of the trans-GZK events. This is 
accomplished without the need to assume exotic astrophysical scenarios. There is no obvious
contradiction with data obtained from FORTE and GLUE, either.\\

\subsubsection{Top-down models}
\vspace{1mm}
Very heavy topological defects (TD) could solve
both the problem of the long trip for protons in the CMB
and the difficulty to accelerate particles to extreme energies~\cite{sigl}.
In the early universe at the time of phase transition from the GUT scale ($M \ge 10^{14}$ GeV) 
TD (monopoles, cosmic strings, necklaces, textures {\em etc.}) could be formed in localized 
regions where mass-energy is trapped. In the decay (or annihilation) of these superheavy
extragalactic relics, ``X-particles'' are formed which decay trough QCD fragmentation jets consisting 
mostly of photons, neutrinos and $\sim 5\%$ nucleons. These nucleons provide the trans-GZK  
events. However, a large fraction of the energy is injected into the extragalactic space
via $\gamma$-s with $E>10^{15}$ eV. 
Thus, if the TD can generate the flux of  EECR, then the EGRET limit (see {\bf 2.4}) is always
exceeded~\cite{protstan}. \\

Another type of top-down scheme uses the possible existence of superheavy dark matter
particles, which could have been created at the end of inflation in the early Universe. Their decay
or annihilition could also produce the trans-GZK spectrum via 
QCD fragmentation~\cite{todra}. As cold dark matter, they could have an overdensity
in the halo of our galaxy by a factor of $10^4$. Their mass must be $ >10^{12}$ GeV and the extreme
energy protons will reach the detectors since they are produced  closeby.
However, there is anisotropy in the arrival direction
because of our position in the galaxy. In addition, the fragmentation process produces
about a factor of 2 more ultra high energy photons than protons. As it was discussed
in Section {\bf 2.2}, the experimental photon fraction is certainly $< 1$.\\

\subsubsection{Stable Light Neutral Particles}
\vspace{1mm}
The idea of light, supersymmetric, stable or long-lived hadrons instead of nucleons
as UHECR was entertained for some time.
A mass larger than the mass of the nucleon ($\approx 2$ GeV) would increase the threshold energy of
the $\pi$ photoproduction  and  push the GZK cutoff to higher energies. \\ 

One hypothetical particle in this class~\cite{farrar} required the existence of a light gluino. 
Precision measurements at LEP and SLC closed the so-called ``light gluino window''~\cite{janot}.
Shadrons with a light sbottom were investigated as well~\cite{sbottom}, motivated by the preliminary result
from the Tevatron indicating an  overproduction of {\em $b\bar{b}$}. Futher data taking, however,
did not confirm the effect~\cite{mary}.\\

The possibility of axion-photon mixing was considered in Ref.~\cite{csaba}. The hypothesis
was that the trans-GZK showers were started by regenerated photons, while the axions were the 
long distance travelers. One may object, however, that according to present observational data
the trans-GZK showers  look hadronic and not photon-induced.

\subsubsection{Lorentz Invariance Violation}
\vspace{1mm} 
The idea of  small violation of Lorentz invariance is not new~\cite{sato}.
 The idea was reconsidered again as a 
possible explanation for the lack of the GZK cutoff in the AGASA and Fly's Eye 
data~\cite{colemanglashow}
as well as a putative absence of high energy photon absorption on the infrared
(IR) background~\cite{protheroemeyer}\footnote{We were informed by Michael Fall
(private communication) that the uncertainties in our knowledge of the IR
 background are considerable.}. A review of the various schemes can be 
found in ref.~\cite{steckeretal}. There is no firmly established scheme of violating 
Lorentz invariance at
present. The most popular scenario invoked in explaining the 
existence of trans-GZK cosmic ray events follows the paradigm laid out in Ref.~\cite{colemanglashow}.
 According to that scheme, Lorentz invariance is
violated by assigning different maximal speeds to different particle species 
(for instance, to electrons and photons), none of them or some of them
being equal to what is known to be the speed of light  {\em in vacuo}.\\

In this manner, the rate of pair production by photons and the rate of
photoproduction by protons can be suppressed by invoking a rather
small amount of Lorentz invariance violation. Conversely, the absence of trans-GZK flux in 
future experiments could place severe constraints on Lorentz invariance violation.
It is important to point out that the same set of assumptions
leads to some rather dramatic changes in the development  of cascade  showers,
see ref.~\cite{vankovstanev}. 
A  violation of Lorentz invariance as an explanation of trans-GZK event cannot be ruled out
at present.\\

\section{Conclusion}
\vspace{2mm}
A plethora of particle physics models was created to explain the trans-GZK events,
which are anomalous according to expectations based on astrophysics and on SM.
Z-bursts, TD  seem to be ruled out already to a large extent
by  the extragalactic diffuse $\gamma$-ray background measurements. 
Superheavy dark matter is fatally wounded according to the available observational data
on the ratio of UHECR photons to protons and on the lack of directional asymmetry. 
Strongly interacting neutrinos and Lorentz invariance violation are still alive. 
The nature and origin of EHECR is still an enigma. If the observational 
data from the next generation of detectors support the existence of trans-GZK events
then the next great leap for particle physics could come from non-accelerator experimental
physics. It is evident that our physics and astrophysics ideas can truly be changed 
by the results coming from the continuing and  the newly commissioned detectors: HiRes,
Telescope Array, P. Auger Observatory, EUSO, OWL, AMANDA, ICECUBE, NESTOR, NEMO, ANTARES,
HESS, VERITAS, MAGIC, GLUE, FORTE etc. 

\section{Acknowledgements}
\vspace{1mm}
We would like to thank V. Berezinsky and W. Burgett for valuable discussions.
Special thanks are due to the organizers of XIII ISVHECRI, in particular to Leo Resvanis
for hosting such an informative and stimulating meeting. We further thank the organizers 
for financial support.

\end{document}